\documentclass[preprint]{aastex}

\begin{document}

\title{A Warp in the LMC Disk?}

\author{K.A.G. Olsen\altaffilmark{1}\affil{National Optical Astronomy Observatory, CTIO, Casilla 603, La Serena, Chile\\ kolsen@noao.edu}}
\altaffiltext{1}{Visiting Astronomer, Cerro Tololo InterAmerican
Observatory, National Optical Astronomy Observatory, which is
operated by Associated Universities for Research in Astronomy, Inc.,
under cooperative agreement with the National Science Foundation}
\author{C. Salyk\altaffilmark{2}\affil{Massachusetts Institute of Technology, Department of Earth and Planetary Science, Building 54, Cambridge, MA, 02139\\ csalyk@mit.edu}}
\altaffiltext{2}{Participant in the 2002 CTIO Research Experiences for Undergraduates program, which is supported by the National Science Foundation}

\begin{abstract}
We present a study of the shape of the Large Magellanic Cloud disk.  We use the brightnesses of core helium-burning red clump stars identified in $V-I,I$ color-magnitude diagrams of 50 randomly selected LMC fields, observed with the CTIO 0.9-m telescope, to measure relative distances to the fields.  Random photometric errors and errors in the calibration are controlled to $\lesssim$1\%.
Following correction for reddening measured through the color of the red clump, we solve for the inclination and position angle of the line of nodes of the tilted plane of the LMC, finding $i=35\fdg8\pm2\fdg4$ and $\theta=145\arcdeg\pm4\arcdeg$.  Our solution requires that we exclude 15 fields in the southwest of the LMC which have red clump magnitudes $\sim$0.1 magnitudes brighter than the fitted plane.  On the basis of these fields, we argue that the LMC disk is warped and twisted, containing features that extend up to 2.5 kpc out of the plane.  We argue that alternative ways of producing red clump stars brighter than expected, such as variations in age and metallicity of the stars, are unlikely to explain our observations.  
\end{abstract}

\section{Introduction}
Among the many virtues of the Large Magellanic Cloud (LMC), one of the
most important is that in spite of its proximity, for most purposes
its stars may be considered to lie at a single distance.  As a result,
the LMC is frequently used in studies of stellar evolution, is of
primary importance in the calibration of the distance scale, and is
 the target of several projects aimed at
detecting microlensing events in the Galactic halo (e.g. Alcock et al.\ 2000, Lasserre et al.\ 2000, Udalski et al.\ 1999, Drake et al.\ 2002).  The
neighboring Small Magellanic Cloud (SMC) has never reached the
same level of popularity, for the interesting reason that it is stretched by
several
kpc along the line of sight (e.g.\ Caldwell \& Coulson 1986, Hatzidimitriou \& Hawkins 1989), the probable result of tidal
interactions with the LMC and the Milky Way.

Probing deeper, we find that the stars of the LMC are not, after all, quite
 at the same distance, because the LMC disk is tilted with respect to the
line of sight.  The LMC's tilt has most recently been measured by van
der Marel \& Cioni (2001), who found its value to be $\sim$35\arcdeg
~using asymptotic giant branch and red giant branch tip (TRGB) stars observed
in the 2MASS survey (Skrutskie 1998).  The effect of the tilt is to produce a modulation
of the apparent luminosity of these tracers with position angle in the
LMC's disk, with an amplitude of $\sim$0.2 magnitude peak-to-trough.
Moreover, upon deprojecting the 2MASS data, van der Marel (2001) found
that the LMC disk is elliptical and has a nonuniform surface density
distribution, indicative of tidal forces acting on the LMC.  Clearly, the LMC
is not slipping through its interactions with the Milky Way and the
SMC completely unscathed.

We are thus led to ask, what further evidence for the LMC-SMC-Milky
Way interaction may we find hidden in the LMC's geometry?  The
question is important for chronicling the interaction history through
numerical models (e.g.\ Gardiner \& Noguchi 1996), and may have implications for
interpreting the microlensing results (e.g.\ Zaritsky \& Lin 1997, Zhao \& Evans 2000).  In this paper, we report the results of our use of the apparent
luminosity of ``red clump'' stars to measure the shape of the LMC
disk.  The red clump, which is composed of the younger, more
metal-rich analogues to the core helium-burning horizontal branch
stars found in globular clusters, has been widely discussed in the
literature for use as a distance indicator (e.g.\ Girardi \& Salaris 2001, GS01; Cole 1998).  While the red clump remains controversial as an absolute distance indicator, by using it to measure {\em relative} distances to LMC fields we avoid the disagreement over its zero point.
Red clump
stars are more numerous than TRGB stars by a factor of $\sim$100 and
have tightly defined colors and magnitudes, making them easily
identifiable in color-magnitude diagrams (CMDs).  In the LMC, they
have a typical surface density of 3.5$\times10^4$ stars per square degree, allowing
us to search for structure with high spatial frequency.  Because they
have well-defined mean colors in addition to magnitudes, we can
measure reddenings with the same tracers used to measure distances,
avoiding the population-dependent reddening effects discussed by
Zaritsky (1999).  The errors associated with using the red clump as a distance indicator are mainly systematic ones.  We discuss their possible effect on our results in section 4.

\section{Observations}
We observed 50 fields in the LMC with the CTIO 0.9-m telescope and
SITe 2K$\times$2K CCD\#3 camera during the nights 2001 November
18--24 (Fig. 1).  We used CTIO's Johnson $V$ and Kron-Cousins $I$ filters.  The fields were randomly selected from a 6\arcdeg$\times$6\arcdeg area of the LMC, with the constraint that they should lie far from regions of recent star formation, where we expect extinction by interstellar dust to be high.  We used gain setting 2, which provides gain of $\sim$1.5 $e^-$/ADU and read
noise of $\sim$3.5 $e^-$.  All of the nights were photometric, while the seeing was
typically 1\arcsec--1\farcs5.  Starting with our first field, our
procedure was to take one pair of $V$ and $I$ exposures in one field
before moving on to the next.  Once all fields had been observed once
in $V$ and $I$, we repeated the sequence twice.  With 250 sec of
exposure in $V$ and 200 sec in $I$, we were able to achieve
S/N$\sim100$ at $V=19,I=18$ while observing $20-25$ fields per night.  In
all, we obtained 3--4 pairs of $VI$ images per field, each observed on
different nights.

We observed Landolt (1992) photometric standards at several times
during each night for calculation of zero points, atmospheric extinctions, and
color corrections.  We also observed two standard star fields located
near the position of the LMC, for which Arlo Landolt kindly provided
standard magnitudes.  These fields were used to check for different
atmospheric extinction at the LMC's position compared to the celestial
equator, for which we found no evidence.

A byproduct of the fast quad amplifier readout used with the SITe
2K$\times$2K CCD is crosstalk between each amplifier and the other
three, which produces faint ghosts in the images.  Our first image
reduction step was to remove these ghosts by subtracting from each CCD
section the proper fraction of the neighboring sections.  Following
this crosstalk correction, we performed the standard image reduction
steps of overscan subtraction, trimming, bias subtraction, and flat
fielding with twilight sky flats, all done with the QUADPROC package
in IRAF.

\section{Analysis}

\subsection{Photometry}
We performed photometry on each of the 304 individual reduced frames
with DAOPHOT (Stetson 1987) and the accompanying profile-fitting program ALLSTAR.
Because finding isolated stars in the crowded LMC fields was
difficult, we selected a large number ($\sim$200) of stars per frame
to define the point-spread function (PSF), allowing an average,
quadratically variable PSF to be constructed.  We corrected the PSF
magnitudes to an aperture of diameter 15\arcsec ~by performing aperture
photometry on the PSF stars after subtracting all other stars, while for
some stars allowing the magnitude to be extrapolated based on smaller
apertures using the aperture growth curve measured using DAOGROW
(Stetson 1990).

\subsection{Photometric calibration}
We measured only aperture magnitudes for the standard star frames,
again using DAOGROW to extrapolate the measurements of some of the
stars to a 15\arcsec ~aperture.  We then calculated nightly zero
points, atmospheric extinction coefficients, and color terms.  After averaging the
seven color term measurements, we recalculated the zero points and
atmospheric extinctions while holding the color terms fixed.  We applied these calibration equations to our LMC
frames using the PSF stars as local standards.  Table 1 shows the calibration coefficients of the transformation equations:
\begin{displaymath}
v = V + A_0 + A_1(V-I) + A_2X
\end{displaymath}
\begin{displaymath}
i = I + B_0 + B_1(V-I) + B_2X
\end{displaymath}
where uppercase refers to standard magnitudes, lowercase to instrumental magnitudes, and $X$ is the airmass.

\subsection{Measurement of red clump stars}
Each frame contains several thousand red clump stars, easily
identified by their location in the CMD.  After selecting stars within
a box encompassing the red clump (Fig.\ 2), we fit a Gaussian profile
combined with a second order polynomial to the $V-I$ and $I$
histograms within the box, taking the Gaussian means as the mean color
and magnitude of the red clump stars.  We used bin sizes corresponding approximately to the precision which we
can measure the means, given the intrinsic $V-I$ and $I$ dispersions
of the histograms and the number of red clump stars per frame.  We
filtered the histograms with a 20--40 bin-wide Savitzky-Golay filter
(Press et al. 1992) before fitting, as we found that our fitting routine behaved more
reliably with this step included.  However, the $\chi^2$ values and
likelihoods of the fits were calculated using the unfiltered
histograms.  On average, we found the fits to the $V-I$ histograms produced $\chi^2/\nu$=1.3 with $\sim$20\% chance of measuring higher $\chi^2$ purely by chance; for the $I$ histograms the numbers were $\chi^2/\nu$=1.04 with $\sim$35\% probability for higher $\chi^2$.

Next, we averaged the 3--4 measurements per field to produce an
average red clump $V-I$ color and $I$ magnitude for each field.
Assuming an intrinsic red clump color of $V-I$=0.92, we calculated
$E(V-I)$ interstellar reddening values for each field and converted these to
$I$-band interstellar extinctions through the equation $A_I = 1.4E(V-I)$ (Schlegel, Finkbeiner, \& Davis 1998).  We
selected the intrinsic color so as to produce a median reddening equal
to that measured by Schlegel et al. (1998) for the LMC, but note that
our choice has no effect on our results since we are using the red
clump only to measure relative distances.

Our formal errors in measuring the mean color and magnitude of the red
clumps are small, $<<$1\%.  We thus expect our distance errors to be
dominated by the
photometric calibration (1\%) and by error in assigning the correct
reddening through the red clump color (2\%).  We expect that variations in the
age and metallicity of the red clump from field to field will be $\sim$2\%; however, we do not include this source in our budget of random distance errors, as it is likely to be correlated with position in the LMC.  We discuss the sources of error in depth in section 4.

\subsection{The LMC's geometry}
For the purposes of this section, we assume that the variation in the
reddening-corrected, mean $I$ magnitudes of the red clump stars in our
fields is produced purely by distance effects, so that we can study
the LMC's geometry.  First, we converted the celestial coordinates of
our fields into planar coordinates following Wesselink (1959).
Second, we transformed the observed $\Delta I=I^{RC}-I^{RC}_\circ$
into distances relative to the center of the LMC, which we take to be
05$^{\rm h}$19$^{\rm m}$38\fs0 $-$69\arcdeg27\arcmin5\farcs2 (2000.0, precessed from 1950.0; de Vaucouleurs
\& Freeman 1973).  We note that our assumed origin has no effect on
our study of the LMC's shape.  Finally, we corrected for the small
geometric projection effect produced by our fixed vantage point.

A weighted least-squares fit of a plane to all of the data yields a
disk inclination angle $i=24\arcdeg\pm2\arcdeg$ and position angle of
the line of nodes $\theta=149\arcdeg\pm6\arcdeg$, where we have used
the convention of $i=0\arcdeg$ for a face-on disk and measured the
position angle as the number of degrees east with respect to north.
This measurement is in good agreement with e.g.\ Caldwell \& Coulson
(1986), who found $i=28\fdg6\pm5\fdg9$ and $\theta=142\fdg4\pm7\fdg7$
from their analysis of individual Cepheid distances.  However, there is a possible problem with the fit, as
$\chi^2$ per degree of freedom is equal to 1.4, with only a 2\% chance of attaining
$\chi^2$ as high as ours through chance alone.

Figure 3 (right panel) shows the LMC disk as viewed edge-on along our measured line
of nodes.  The tilt of the disk is strikingly well defined,
particularly when compared to Figure 6 of Caldwell \& Coulson (1986).
However, the fields at the southwestern edge (open circles) fall systematically low compared to the fitted plane, by $\sim$0.1 magnitudes in $I$ or 2.5 kpc in
distance.  If we remove these 15 fields (labelled by diamonds in the left panel of Figure 3) from the fit, we find instead
$i=35\fdg8\pm2\fdg4$ and $\theta=145\arcdeg\pm4\arcdeg$.  This
inclination value is in excellent agreement with the value
$i=34\fdg7\pm6\fdg2$ measured by van der Marel \& Cioni (2001),
although we disagree at the $\sim2\sigma$ level with their position
angle $\theta=122\fdg5\pm8\fdg3$.  The statistical likelihood of the
fit is now excellent, with a 96\% chance of producing $\chi^2$ values
higher than our measured 0.62 per degree of freedom.  Indeed, the
high likelihood of the fit suggests that we have overestimated the
size of our error bars.  

Fitting a plane to only the 15 southwestern fields which we removed above, we find $i=14\arcdeg\pm9\arcdeg$ and $\theta=280\arcdeg\pm47\arcdeg$, with $\chi^2/\nu=0.5$ and 94\% probability of producing higher $\chi^2$ by chance.  Thus, our data imply that the LMC disk is warped and that its line of nodes is twisted at the southwestern edge.

\section{Discussion}
As shown by fitting planes to both the 15 southwestern and 35 remaining fields, our analysis suggests that the disk of the LMC is warped and twisted.  This result can perhaps explain the different tilts measured by Caldwell \& Coulson (1986) and van der Marel \& Cioni (2001) as being caused by differing contributions of the warp to their datasets.  The area sampled by our observations is similar to that studied by Caldwell \& Coulson (1986), so that we expect similar results if we include all of our fields in the fit.  Moreover, van der Marel \& Cioni (2001) find that the inclination of the LMC drops by $\sim$10 degrees outside a radius of $\sim$4\fdg5 from the center.  Although our study suggests that the LMC's warp begins at a radius $\sim$3\arcdeg from the center, van der Marel \& Cioni (2001) surveyed a much larger area than we have, making a direct comparison with their study difficult.

Our finding that the LMC is warped depends, however, on our assumption that distance is the dominant effect producing the observed variation in red clump luminosity.  In the following sections, we discuss the possible sources of systematic error.
\subsection{Uncertainty in the calibration}
The stability of the derived photometric zero points and atmospheric extinction
terms suggests that the photometric calibration is accurate to
$\lesssim$1\% (Table 1).  A further check against systematic errors is provided
by the fact that we observed every field on three different nights.
In Figure 4, we plot $V-I$ and $I$ of the red clump for the 152 pairs
of observations, having subtracted the mean clump $V-I$ and $I$ for
each field.  The nightly drift and scatter is smaller than 1\%, ruling out
calibration uncertainty as the cause of the apparent warp.

\subsection{Reddening}
If we have overestimated the reddening $E(V-I)$ in the 15 southwestern
fields by $\sim$0.07 magnitudes compared to the other fields, then
this could explain the effect we see in Figure 3.  Figure 5, where we
plot $A_I$ as derived from the color of the red clump versus the
departure of the fields in $I$ from the plane with $i=35\arcdeg$, is
cause for suspicion.  The southwestern fields have red clumps with systematically higher interstellar extinction compared to the other
fields.  Moreover, the magnitude deviation from the plane is correlated with $A_I$, such that if we have overestimated the reddening in the southwestern fields by more than a factor of two, then reddening could explain the supposed warp.

We checked whether we have miscalculated the reddening by looking for
differences in other evolutionary features in the color-magnitude
diagrams of the fields.  If the reddening values are wrong, the error
should manifest itself as a difference in color between the red clump
and the red giant branch and main sequence in the southwestern fields.
After correcting for the reddening and red clump luminosity
differences, we merged the photometry for the southwestern fields and
the remaining fields separately.  Figure 6 shows the difference of the two-dimensional histograms of the CMDs of the merged southwestern and other fields, after applying various shifts in $V-I$ to the southwestern fields.  The overlay of the colors of
the main sequence and red giant branch allow a $\sim$0.01--0.02
magnitude error in $E(V-I)$, but not a difference as large as 0.07
magnitudes.  We therefore rule out large reddening errors as the cause
of the warp.

If we have properly accounted for interstellar correction, then why is there a correlation between the magnitude deviation from the LMC plane and $A_I$?  Figure 7 offers a plausible explanation.  The COBE/DIRBE--IRAS/ISSA dust map (Schlegel et al.\ 1998) shows that the southwestern corner of the LMC is associated with a region of higher diffuse extinction, presumably due to Galactic foreground dust, having $A_I\sim0.3$.  For comparison, we measured $A_I\sim0.25$ at the LMC's southwest edge.

\subsection{Age and metallicity}
GS01 and Cole (1998) have discussed the strong
effect that age and metallicity have on the luminosity of red clump
stars.  In short, metal-poor clumps are brighter than metal-rich ones,
and younger clumps brighter than older ones.  Assuming a mean red
clump age of $\sim$4 Gyr, appropriate for the LMC (GS01), and a mean
metallicity of [Fe/H]$\sim-0.6$ (Pagel \& Tautvaisiene 1998), $M_I$ of
the red clump may be made brighter by 0.1 magnitude by either lowering
[Fe/H] by $\sim$1 dex or by lowering the age by $\sim$2 Gyr; $M_I$
could be made fainter by 0.1 magnitude by raising [Fe/H] by 0.3 dex or
by increasing the age by $\sim$3 Gyr (GS01).  Thus,
we need to consider that the supposed warp may be produced by the
southwestern fields being more metal-poor and/or younger than the rest
of the LMC disk.

There are a number of arguments against this hypothesis.  First, the appearance of the CMDs appear to rule out large bulk
differences in age or metallicity between the southwestern and the
remaining fields.  Figure 8 shows the combined CMDs plotted against
Girardi et al.\ (2000) isochrones with ages of 2 and 4 Gyr and [Fe/H]
equal to $-0.6$ and $-1.6$.  The similar slopes of the red giant
branches in the southwestern fields compared to the other fields suggest
that the ages and metallicities of the red clump stars are also
similar.  

Second, the variation due to age and metallicity of the $I$ luminosity of the red clump is predicted to be small across the LMC.  GS01, using the LMC Bar and outer disk star formation histories derived by Holtzman et al.\ (1999) coupled with the age-metallicity relation predicted by Pagel \& Tautvaisiene (1998), calculated that the Bar red clump should be only 0.01 $I$ magnitudes brighter than the outer disk red clump.  We derive a similar result by feeding the star formation histories derived by Smecker-Hane et al. (2002) for the LMC Bar and disk through the GS01 models.  We find that the LMC Bar should have a red clump brighter than the disk by $\lesssim$0.03 magnitudes, despite the Bar having a $\sim$2 Gyr younger age. For comparison, Smecker-Hane et al. (2002) find that the red clump in their Bar field is 0.02 magnitudes brighter in $I$ than in their disk field.  After correcting for the 0.04 magnitude larger $A_I$ in their Bar field and for our solution of the LMC's tilt, which places their Bar field 0.02 magnitudes closer than their disk field, we find that their Bar red clump is 0.04 magnitudes more luminous than their disk red clump.  This measurement is in good agreement with the theoretical prediction.  Thus, both theory and observations show that the 
mean properties of field red clump
stars are less sensitive to age and metallicity effects than are
single-age, single-metallicity populations.  This is to be expected from the composite nature of stars in the field.

Third, differential rotation in the LMC shreds unbound coherent
structure on the orbital timescale (270 Myr at $R$=3 kpc, adopting $v_{\rm rot}=70$ km s$^{-1}$; Alves \& Nelson 2000), while the bulk of the red clump stars in the LMC are at least
$\sim$4 Gyr old (GS01).  We thus do not expect that an 
area of the
LMC $\sim$2 kpc in size would retain a record of its locally produced age and
metallicity distribution, distinct from fields at similar radii.  A warp, on the other hand, would be a
non-equilibrium, recently formed structure not yet mixed into the surroundings.

Finally, explaining the supposed warp through population effects would
require an age or metallicity gradient that is inconsistent with observed trends.  The general trend in the LMC is for the younger clusters to lie preferentially towards the inner regions (Bica et al.\ 1996), in the oppposite sense from that required here.  The required metallicity gradient ($-1.0$ dex over a stretch of $\sim$2 kpc), while operating in the right sense, is much steeper than the weak radial gradient observed (Kontizas et al.\ 1993).  

In summary, we think that the simplest explanation for the unusually bright
red clump luminosities seen in our southwestern fields is that the LMC
disk is warped, rather than that the southwest of the LMC is more metal-poor or younger than the rest of the disk.  However, a detailed solution of the star 
formation and chemical enrichment histories implied by our fields is needed to answer the question definitively.

\section{Conclusions and future work}
We have studied the shape of the LMC by measuring the variation in
luminosity of red clump stars across 50 13\arcmin$\times$13\arcmin
~fields distributed over a 6\arcdeg$\times$6\arcdeg area.  Our observations were
taken with the CTIO 0.9-m telescope CCD camera using $V$ and $I$
filters.  We clearly detect the tilt of the LMC disk through a
decrease in brightness of the red clump stars along the LMC's NE-SW axis.
Our measured tilt of $35\fdg8\pm2\fdg4$ is in excellent agreement with
the recent measurement of van der Marel \& Cioni (2001), {\em as long
as} we exclude 15 fields in the southwest which appear to lie out of
the plane of the disk.  Including {\em all} of the fields, we measure a
lower inclination of $24\arcdeg\pm2\arcdeg$.  This value is in good
agreement with that measured by Caldwell \& Coulson (1986); however,
the fitted plane is of lower statistical significance.  

We believe that we have detected a warp in the LMC disk which causes a
region $\sim$2 kpc wide in the southwest to lie out of the plane by
$\sim$2.5 kpc.  If it exists, this out-of-plane feature is likely a
non-virialized structure.  How and when was it produced?  The location
of the warp points a finger at the SMC.  According to the model by
Gardiner \& Noguchi (1996), the LMC and SMC endured a close encounter $\sim$200
Myr ago which drew out the material that new occupies the inter-Cloud
region.  It is possible that this interaction also altered the shape
of the LMC disk.  However, the position of the warp also aligns it
with the major axis of the LMC's elliptical disk (van der Marel 2001), the shape of which
van der Marel attributes to the tidal influence of the Milky
Way.  Perhaps our suggested warp is an additional response to the
Milky Way's tidal force?  Weinberg (1998, 2000) studied the mutual
influence of the LMC and Milky Way through numerical simulations.
Although Weinberg found that the long-term effect on the LMC's
structure is simply heating of the disk, transient structures might
also be produced.

Further exploration of the structure of the LMC could determine
whether the warp is reflected in northeastern fields more distant than those observed here.  The
determination of the star formation and chemical enrichment history in
the LMC's southwest would establish with certainty the contribution of
age and metallicity effects to the observed red clump luminosities.
In this study we avoided the LMC's inner regions, including the Bar,
as these are heavily crowded.  Measurement of the LMC's shape and
thickness in these regions is of strong interest for understanding the
contribution of LMC self-lensing to the observed microlensing events
(Zhao \& Evans 2000).  While our results bear no consequence for the
microlensing studies, they do add to the growing body of evidence
suggesting that the LMC is not a simple flattened disk.  A study of
the red clump in the LMC Bar, combined with the continued monitoring
of the LMC for microlensing events by the SuperMACHO project (Drake et al.\ 2002), could provide very interesting results.

\acknowledgements
This work was carried out as part of the 2002 Research Experiences for
Undergraduates (REU) program at CTIO, which is supported by the National
Science Foundation (NSF).  CTIO is part of the National Optical
Astronomy Observatory, which is operated by the Association of
Universities for Research in Astronomy (AURA), Inc., under cooperative
agreement with the NSF.  We acknowledge useful comments from the anonymous
referee.

\newpage

\begin{figure}
\plotone{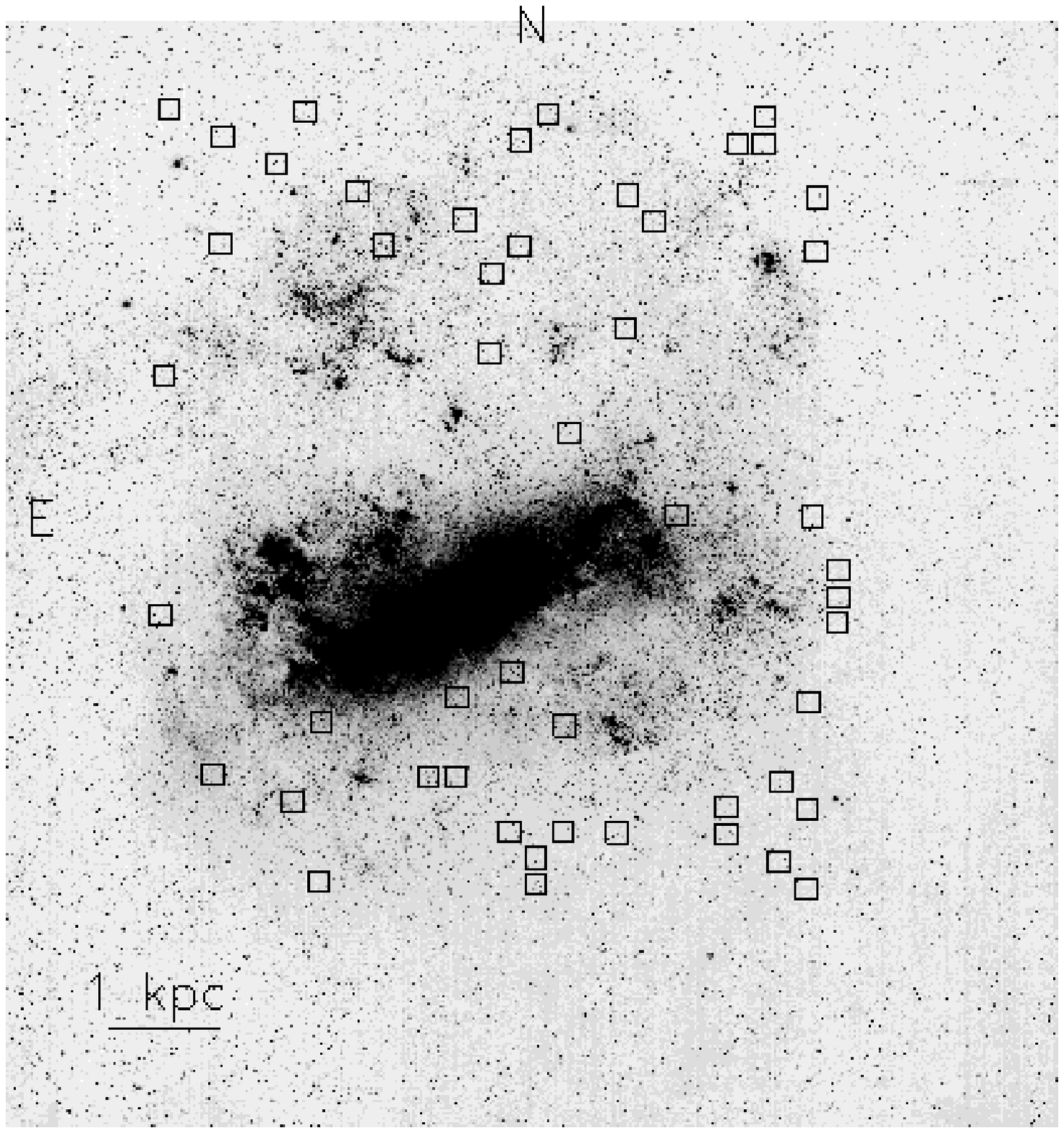}
\caption{The locations of the 50 fields observed with the CTIO 0.9-m telescope in $V$ and $I$ are shown.  The sizes of the boxes are equal to the field size of the CCD camera.  North is up and east is to the left.}
\end{figure}
\begin{figure}
\plotone{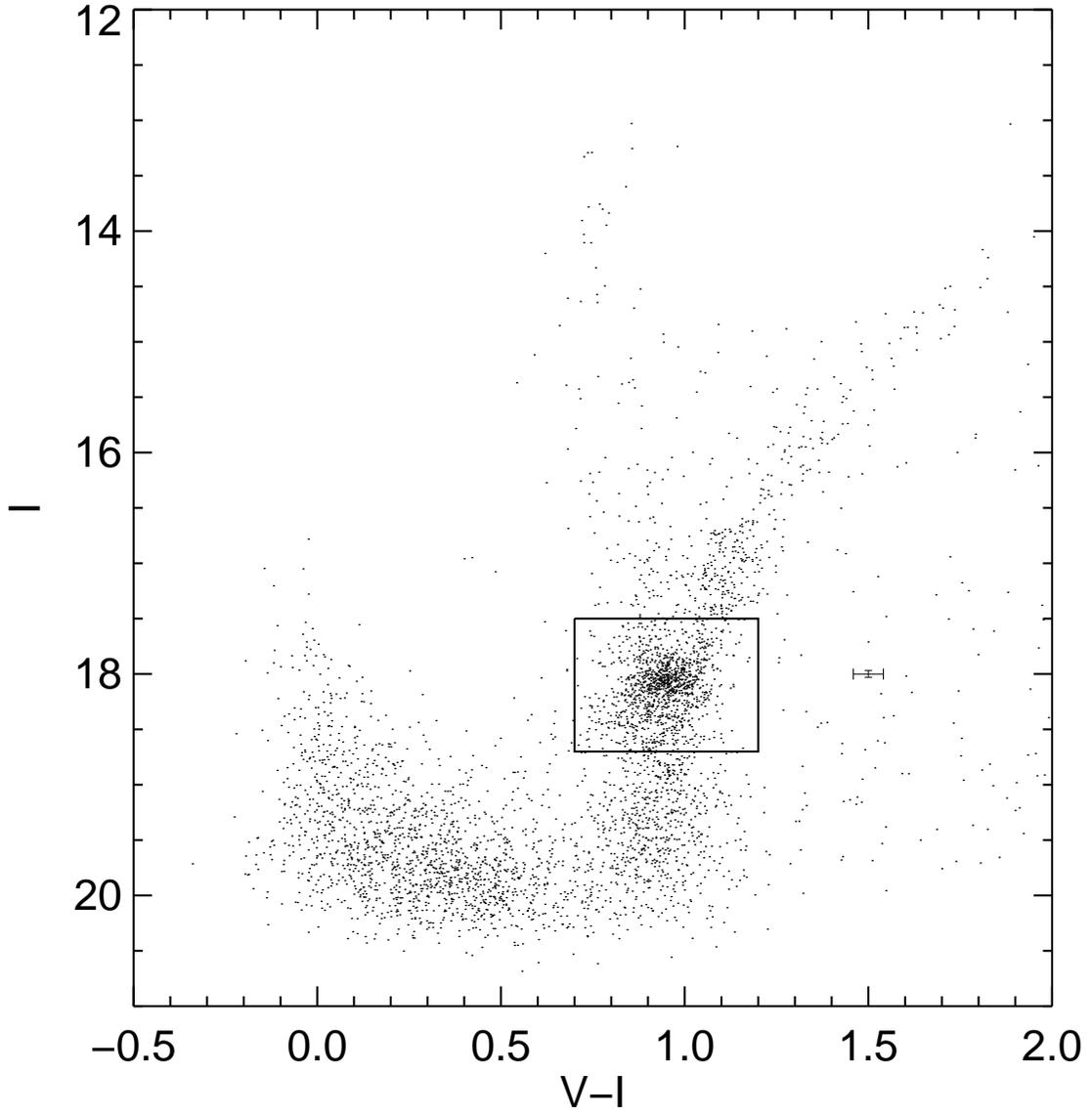}
\caption{A typical LMC color-magnitude diagram.  The box used to select red clump stars is shown.  The error bar indicates the average photometric error in $V-I$ and $I$ as measured by DAOPHOT.}
\end{figure}
\begin{figure}
\plottwo{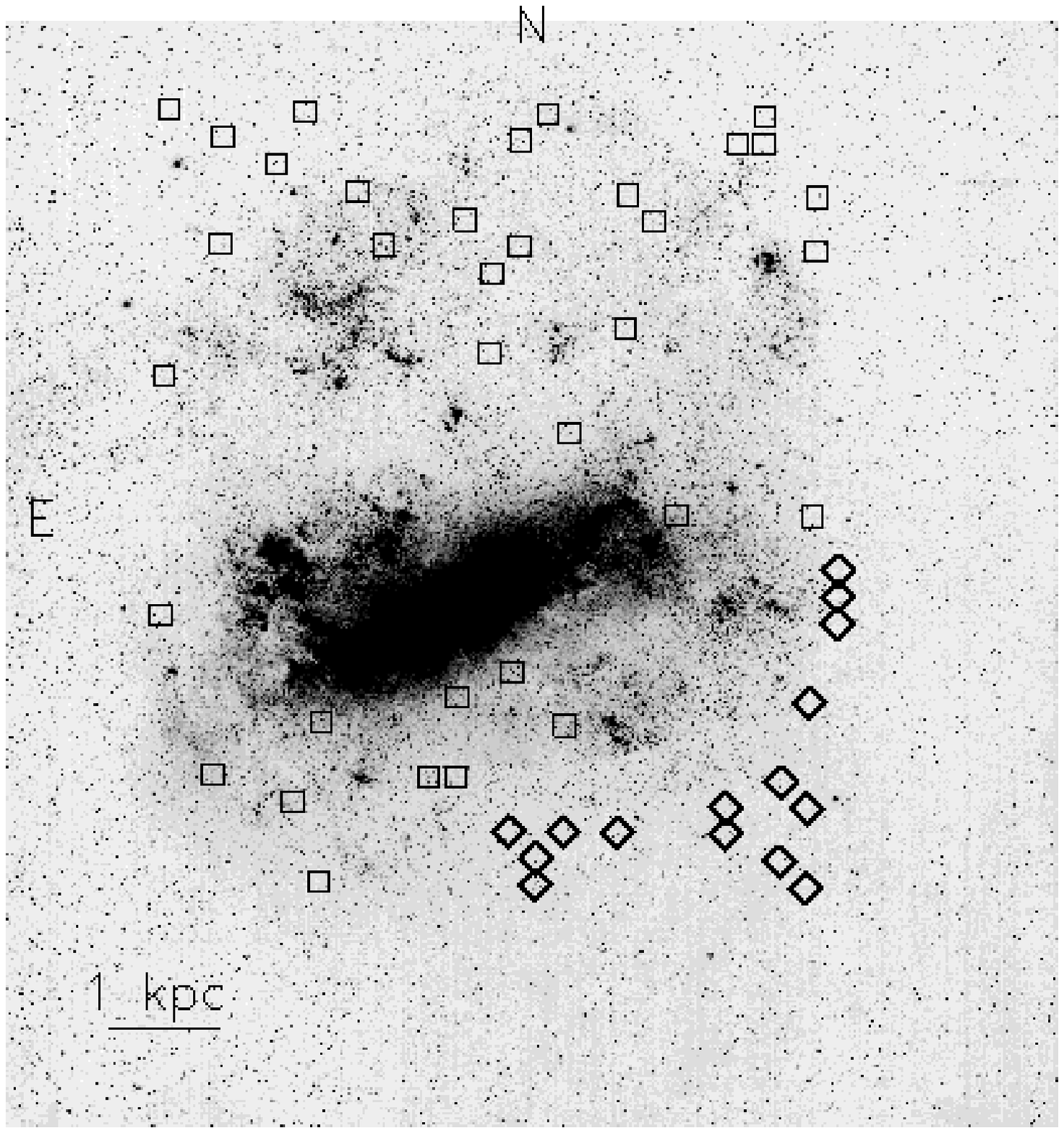}{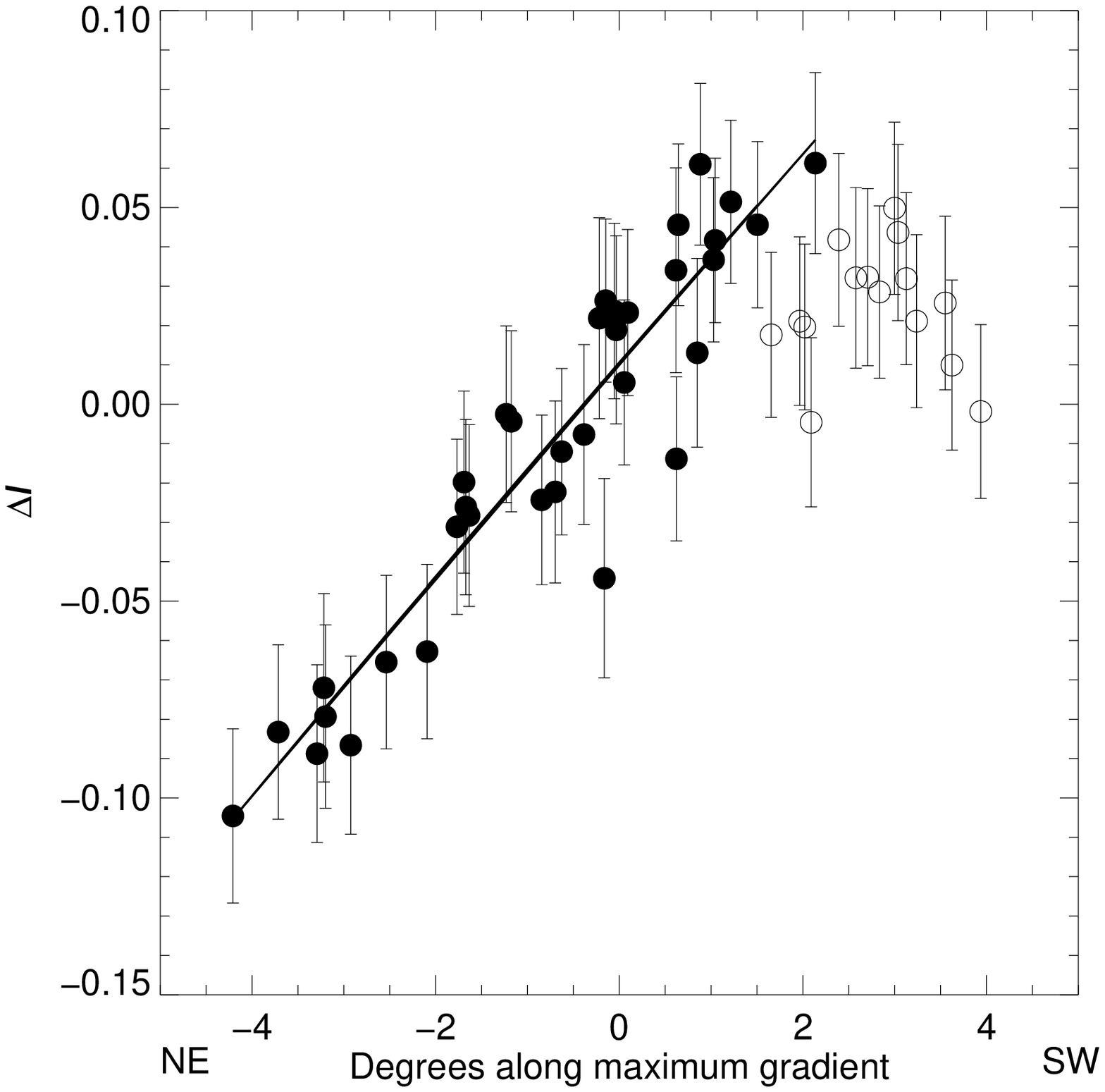}
\caption{The LMC viewed edge-on.  {\it Left}: the locations of the 50 fields shown in Figure 1 are reproduced here.  The 15 fields which appear to lie out of the plane of the LMC are labelled by diamonds, the remaining 35 by squares.  {\it Right}:
the mean-subtracted $I$ magnitudes of red clump stars are plotted as a function of position along the line of maximum gradient of the plane (solid and open circles).  The magnitudes have been corrected for interstellar extinction.  The fitted plane is indicated by the solid line; open circles designate the 15 southwestern fields which were excluded from the fit.  The inclination of the plane is $i=35\fdg8\pm2\fdg4$ and the position angle of the lines of nodes is $\theta=145\arcdeg\pm4\arcdeg$.  Fitting a plane only to the open circles, we find $i=14\arcdeg\pm9\arcdeg$ and $\theta=280\arcdeg\pm47\arcdeg$.
}
\end{figure}
\begin{figure}
\plotone{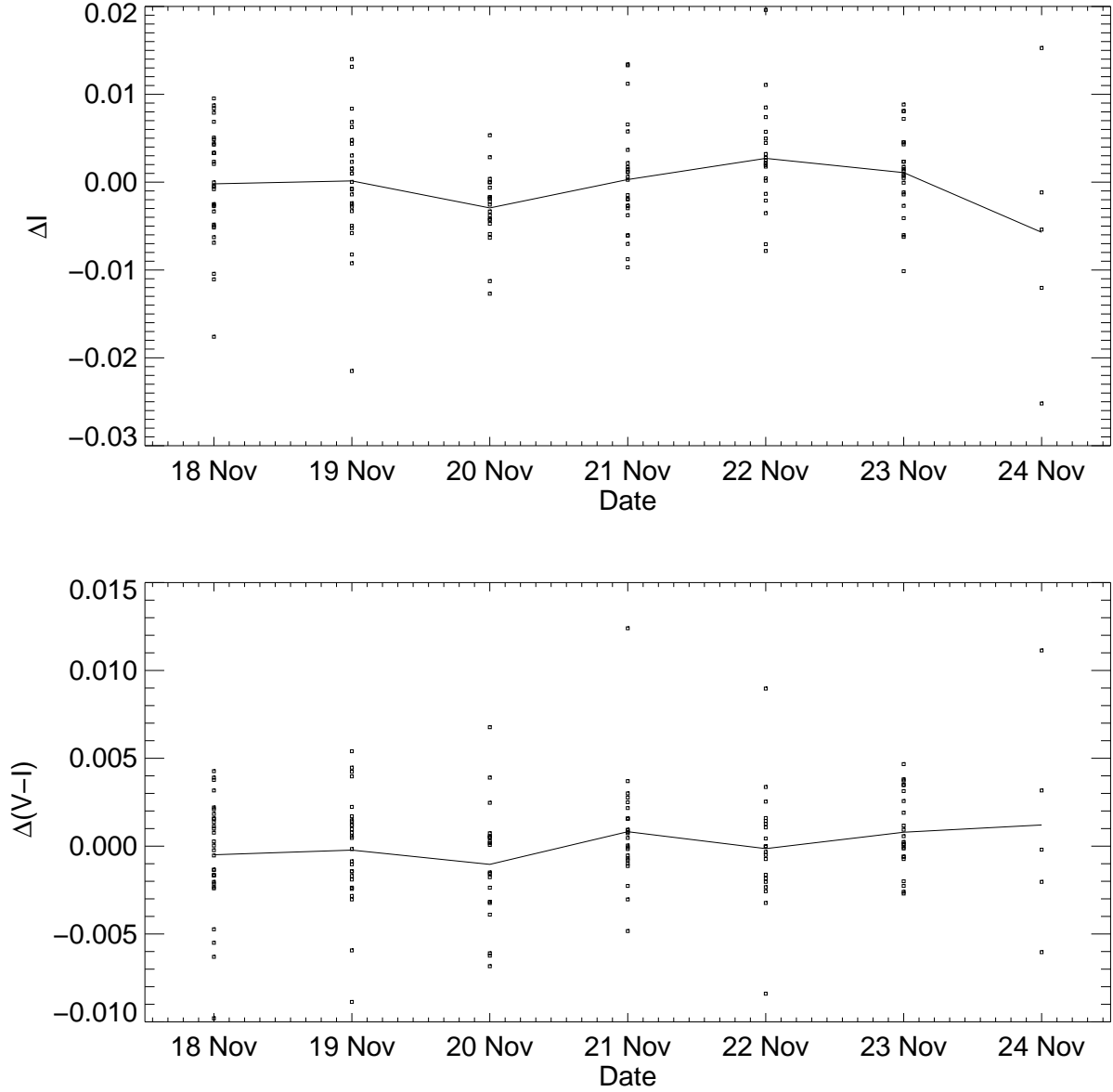}
\caption{Nightly drift in the photometry.  The plots show the individually measured red clump $I$ magnitudes and $V-I$ colors after subtracting the mean of the 3--4 measurements for each field.  The lines indicate the nightly averages, which demonstrate that our photometric calibration uncertainties are $<$1\%.}
\end{figure}
\begin{figure}
\plotone{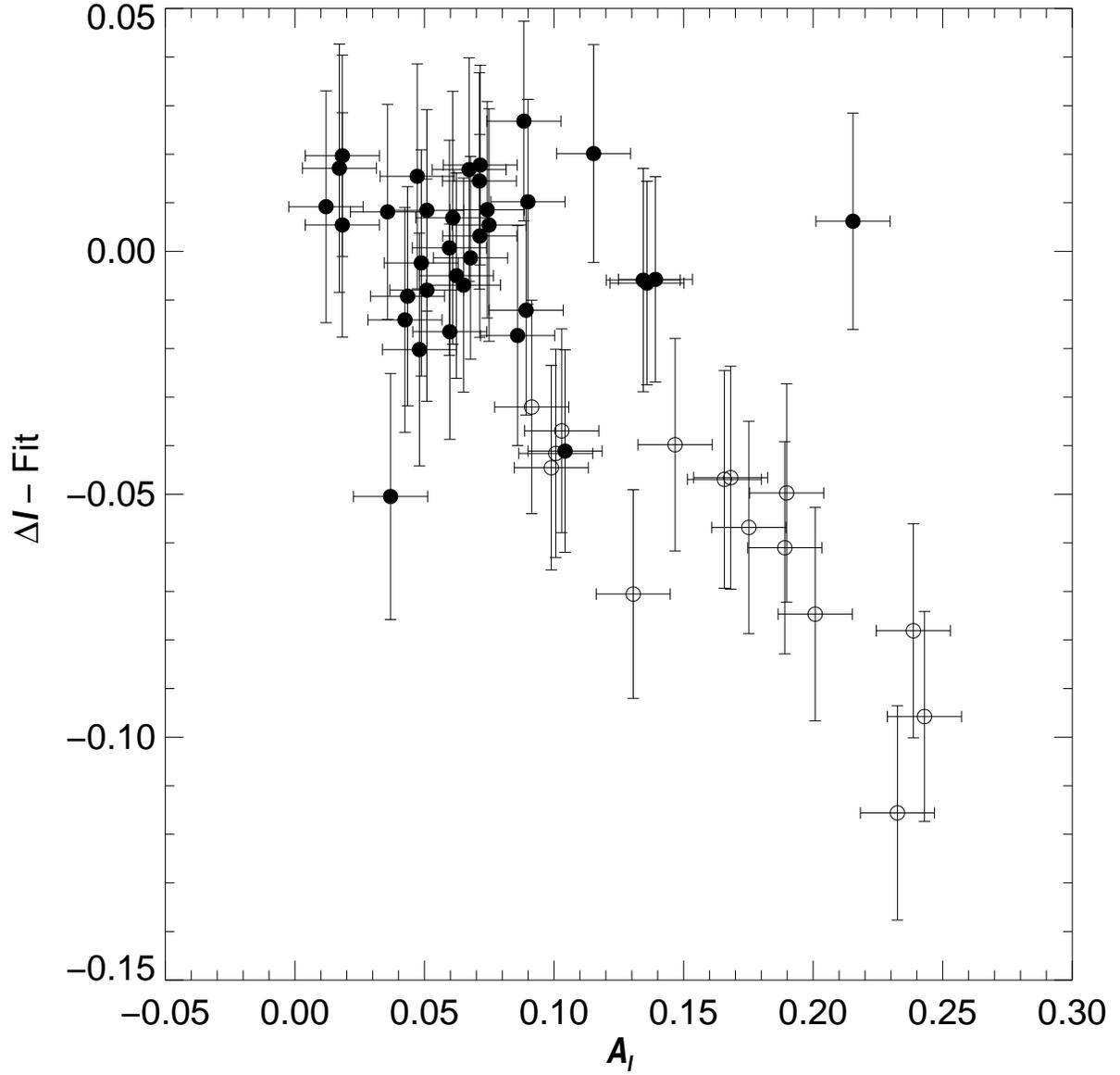}
\caption{The red clump luminosities relative to the fitted plane are plotted against the interstellar extinction values $A_I$, as derived from the color of the red clump.  Open circles are the 15 southwestern fields, filled circles the remaining fields.}
\end{figure}
\begin{figure}
\plotone{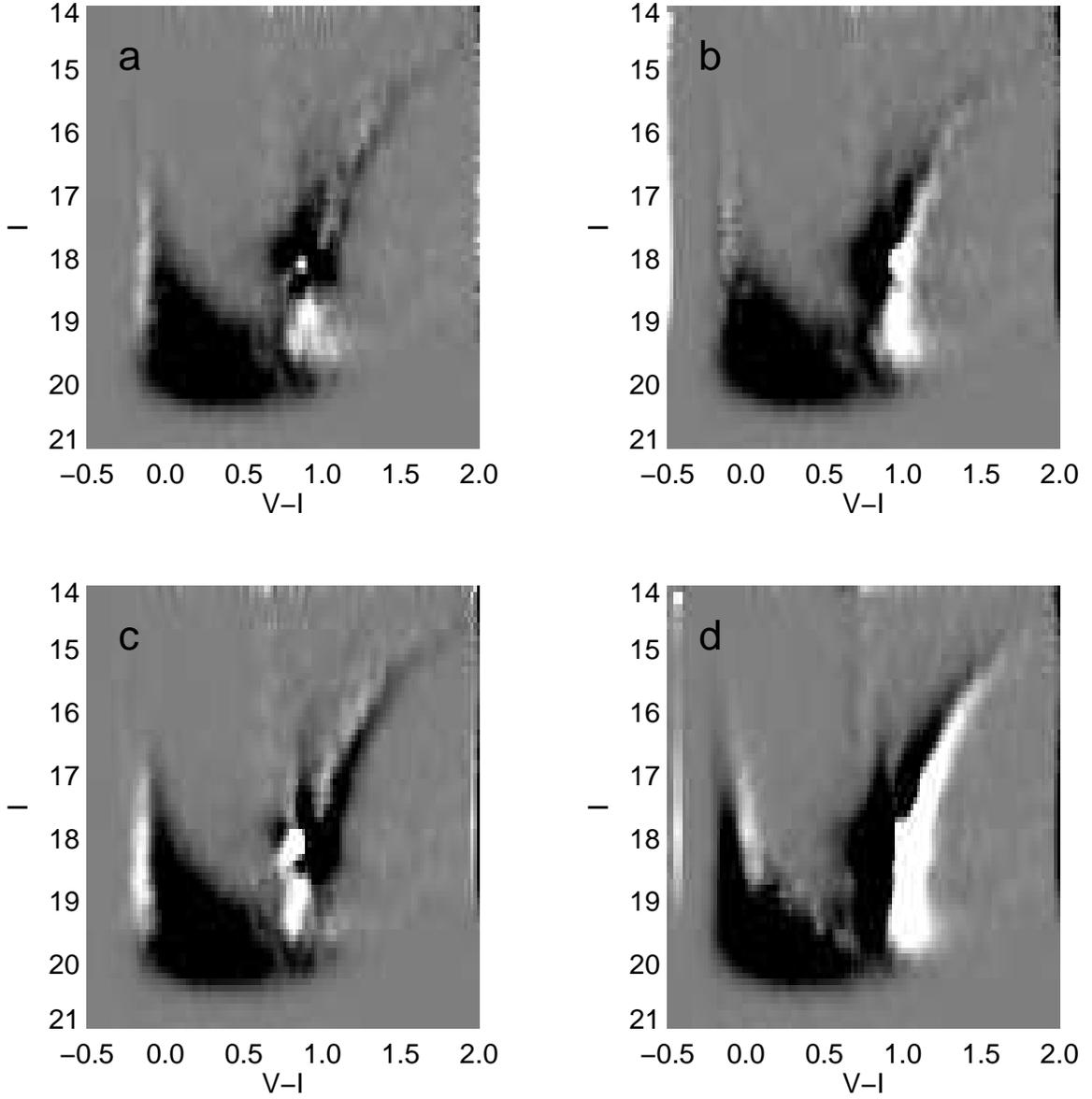}
\caption{The combined CMD of the southwestern fields, expressed as a Hess diagram, are normalized and subtracted with the combined, normalized CMD of the remaining fields.  The CMDs have been shifted so as to line up at the red clump.  White indicates regions where the southwestern fields are high with respect to the other fields; black where the southwestern fields are low.  The panels show the result of applying various $V-I$ shifts to the southwestern CMD: a) no extra shift applied, b) southwestern CMD shifted by 0.02 mags, c) southwestern CMD shifted by -0.02 mags, and d) southwestern CMD shifted by 0.08 mags.  From the overlap of the red giant branches and upper main sequences, we estimate that our reddening uncertainities are $<$2\%.}
\end{figure}
\begin{figure}
\epsscale{0.95}
\plotone{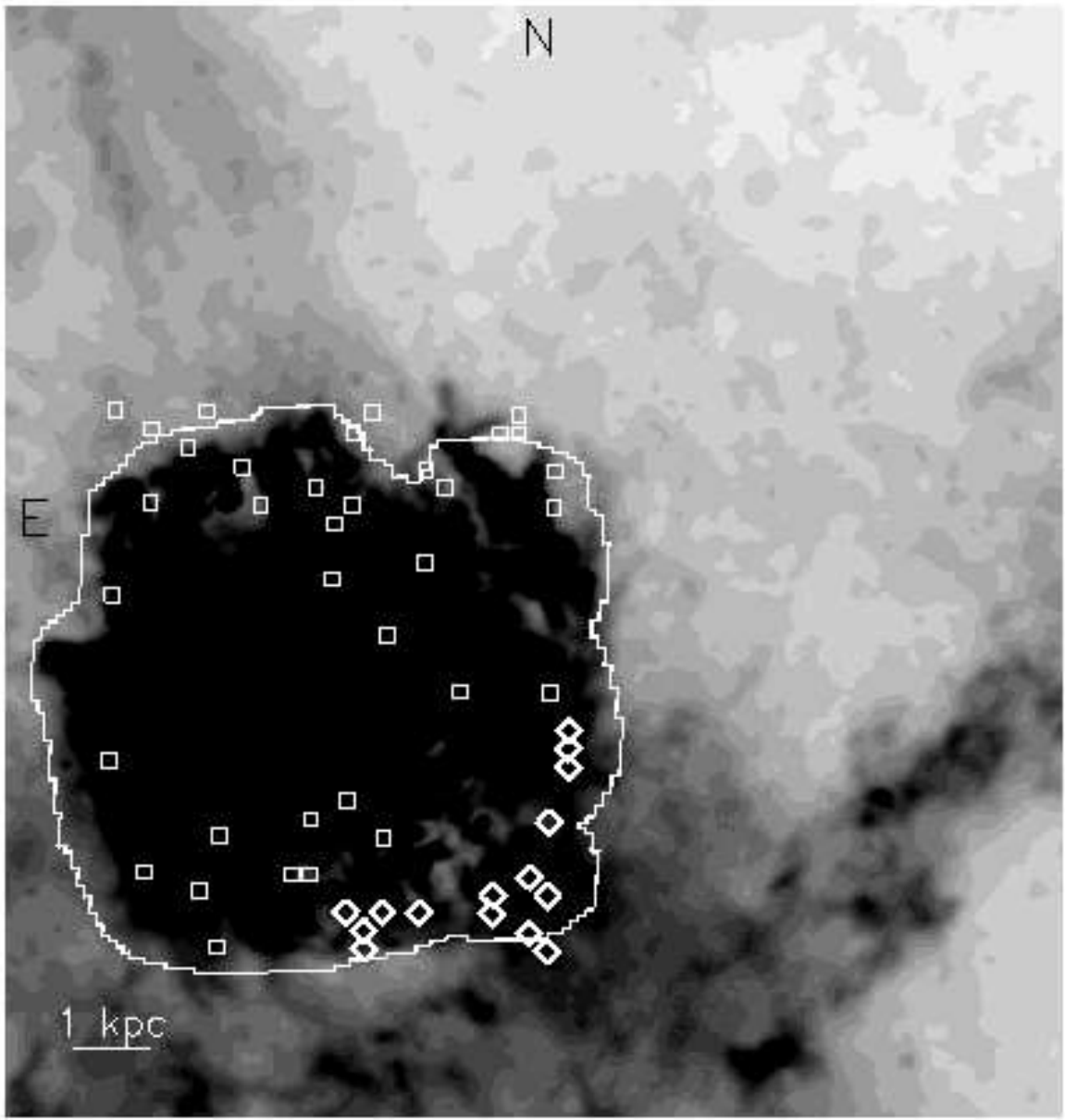}
\caption{Extinction by dust in the vicinity of the LMC.  The grayscale shows a portion of the Schlegel et al.\ (1998) interstellar extinction map of the South Galactic Pole, which was derived from COBE/DIRBE and IRAS/ISSA observations.  The map has been scaled to units of $A_I$.  Darker shades represent higher extinction.  Measurements within the outlined region, which are dominated by LMC emission, are considered inaccurate because the temperature structure is unresolved by DIRBE (Schlegel et al.\ 1998).  The southwestern edge of the LMC extends into a region of enhanced diffuse interstellar extinction, providing a likely explanation for the higher extinction seen in the fields marked by diamonds.}
\end{figure}
\begin{figure}
\epsscale{0.55}
\plotone{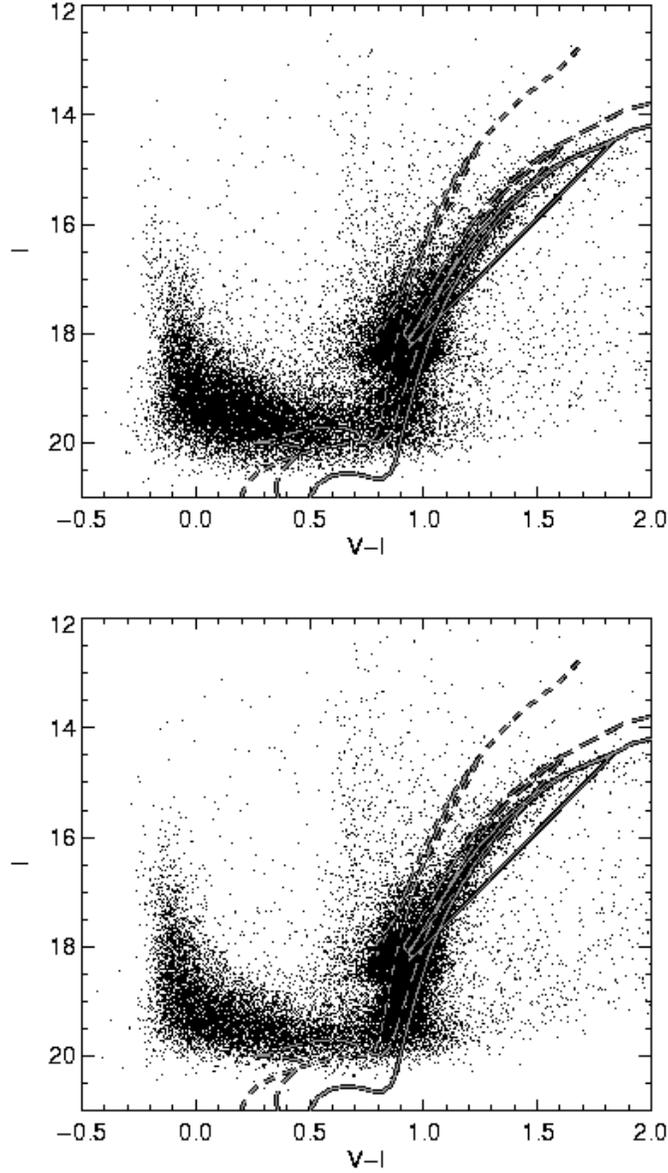}
\caption{{\it Top}: Combined CMD of 35 fields in the LMC.  Only 30000 stars are shown for clarity.  The individual CMDs have been corrected for reddening and shifted in $I$ so as to correct for the tilt of the LMC disk. Solid line: 4 Gyr, [Fe/H]=-0.6 isochrone freom Girardi et al.\ (2000).  Long dashed line: 2 Gyr, [Fe/H]=-0.6 ischrone.  Short-dashed line: 4 Gyr, [Fe/H]=-1.6 isochrone.  {\it Bottom:} As above, but for the combined CMD of the 15 southwestern fields which appear to be participating in the LMC's warp.  The similarity of the CMDs suggest similar stellar populations in the southwestern and remaining fields.
}
\end{figure}

\newpage
\begin{deluxetable}{lccc}
\tablecaption{Photometric calibration coefficients \label{tbl-1}}
\tablewidth{0pt}
\tablehead{
\colhead{Date} &
\colhead{$A_0$} & \colhead{$A_1$} & \colhead{$A_2$} \\
}
\startdata
11/18/01  & 2.124(1) & -0.019(2) & 0.129(4)\\
11/19     & 2.126(1)  &-0.019(2) &0.124(3)   \\
11/20     & 2.135(1)  &-0.019(2) &0.122(5)   \\
11/21     & 2.138(1)  &-0.019(2) &0.123(4)   \\
11/22     & 2.136(1)  &-0.019(2) &0.129(4)   \\
11/23     & 2.119(1)  &-0.019(2) &0.120(4)   \\
11/24     & 2.138(1)  &-0.019(2) &0.124(6)   \\
\tableline
 &  $B_0$ &  $B_1$ &  $B_2$ \\
\tableline
11/18  &2.961(1) & -0.011(2) & 0.068(6) \\
11/19     &2.959(1)   &-0.011(2) & 0.060(4)\\
11/20     &2.963(2)   &-0.011(2) & 0.043(7)\\
11/21     &2.970(1)   &-0.011(2) & 0.038(4)\\
11/22     &2.958(1)   &-0.011(2) & 0.057(4)\\
11/23     &2.966(1)   &-0.011(2) & 0.049(5)\\
11/24     & 2.974(1)  &-0.011(2) & 0.066(8)\\

\enddata
\end{deluxetable}

\end{document}